\documentclass[conference]{IEEEtran}
\IEEEoverridecommandlockouts
\usepackage{algorithmic}
\usepackage{algorithm}
\usepackage{array}
\usepackage[caption=false,font=normalsize,labelfont=sf,textfont=sf]{subfig}
\usepackage{textcomp}
\usepackage{stfloats}
\usepackage{url}
\usepackage{verbatim}
\usepackage{graphicx}
\usepackage{cite}
\usepackage{amsmath,amssymb,amsfonts,amsthm,mathtools}
\usepackage{xcolor}
\usepackage{blindtext}
\usepackage{lipsum}
\usepackage{subfig}
\usepackage{cuted}
\usepackage{physics}
\usepackage[long]{optidef}
\usepackage{multirow}
\usepackage{steinmetz}
\usepackage{bigints}

\newcommand{\indep}{\rotatebox[origin=c]{90}{$\models$}}

\def\BibTeX{{\rm B\kern-.05em{\sc i\kern-.025em b}\kern-.08em
    T\kern-.1667em\lower.7ex\hbox{E}\kern-.125emX}}
\begin{document}

\title{Epistemology-Inspired Bayesian Games for Distributed IoT Uplink Power Control\vspace{-5mm}}
\author{\IEEEauthorblockN{Nirmal D. Wickramasinghe\IEEEauthorrefmark{1}, John Dooley\IEEEauthorrefmark{1}, Dirk Pesch\IEEEauthorrefmark{2}, and Indrakshi Dey\IEEEauthorrefmark{3}}
\IEEEauthorblockA{\IEEEauthorrefmark{1}Department of Electronic Engineering, Maynooth University, Ireland
\\\IEEEauthorrefmark{2}School of Computer Science and Information Technology, University College Cork, Ireland
\\\IEEEauthorrefmark{3}Walton Institute, South East Technological University, Waterford, Ireland
\\\ nirmal.wickramasinghe.2023@mumail.ie,\,john.dooley@mu.ie,\,d.pesch@cs.ucc.ie,\,indrakshi.dey@waltoninstitute.ie\vspace{-5mm}}

\thanks{This work is supported in part by Taighde Éireann - Research Ireland under Grant 13/RC/2077\_P2, by the EU MSCA Project ``COALESCE'' under Grant Number 101130739.}}

\maketitle

\vspace{-10mm}
\begin{abstract}
Massive number of simultaneous Internet of Things (IoT) uplinks strain gateways with interference and energy limits, yet devices often lack neighbors’ Channel State Information (CSI) and cannot sustain centralized Mobile Edge Computing (MEC) or heavy Machine Learning (ML) coordination. Classical Bayesian solvers help with uncertainty but become intractable as users and strategies grow, making lightweight, distributed control essential. In this paper, we introduce the first-ever, novel epistemic Bayesian game for uplink power control under incomplete CSI that operates while suppressing interference among multiple uplink channels from distributed IoT devices firing at the same time. Nodes run inter-/intra-epistemic belief updates over opponents’ strategies, replacing exhaustive expected-utility tables with conditional belief hierarchies. Using an exponential–Gamma SINR model and higher-order utility moments (variance, skewness, kurtosis), the scheme remains computationally lean with a single-round upper bound of $O\!\left(N^{2} S^{2N}\right)$. Precise power control and stronger coverage amid realistic interference: with channel magnitude equal to $1$ and a signal-to-interference-plus-noise ratio (SINR) threshold of $-18$ dB, coverage reaches approximately $60\%$ at approximately $55\%$ of the maximum transmit power; mid-rate devices with a threshold of $-27$ dB achieve full coverage with less than $0.1\%$ of the maximum transmit power. Under $80\%$ interference, a fourth-moment policy cuts average power from approximately $52\%$ to approximately $20\%$ of the maximum transmit power with comparable outage, outperforming expectation-only baselines. These results highlight a principled, computationally lean path to optimal power allocation and higher network coverage under real-world uncertainty within dense, distributed IoT networks.
\end{abstract}

\section{Introduction} \label{Sec: 1_introduction}

\IEEEPARstart{M}{assive} cyber–physical IoT deployments now comprise billions of sensing, actuation, and communication devices that are battery-powered and computationally constrained. These nodes contend for scarce spectrum and energy, often without reliable channel or network-side metrics, making energy-efficient uplink control under incomplete information a first-order challenge for sustainable operation. Gateway-centric schemes can coordinate access and power but frequently incur prohibitive control overheads. For example, pairwise end-device pairing with resource blocks in LoRa uplinks improves efficiency yet depends on gateway scheduling \cite{RA_centralized_LORA}; more broadly, fog-cloud-layer solutions must collect unaware data vectors and exchange acknowledgments, introducing idle slots and added signaling \cite{Survey_RA_Fog_cloud}.
Distributed resource allocation alleviates some of these issues \cite{RA_distributed_IoT}, and iterative sealed-bid auctions provide federated protocols under incomplete information \cite{mSAA_RA_IoT_my}. However, cluster-based computation can impose extensive inter-node message exchange and adapts poorly to fast system dynamics, while deep learning and neural models require prior training and storage that exceed typical IoT device budgets \cite{RA_IoT_AI}.

Game-theoretic models capture strategic interactions among competing devices \cite{Game_theory_IoT_RA}. Bayesian formulations, address uncertainty but suffer exponential growth in complexity with the number of users, degree of unawareness, and strategy cardinality, which limits scalability in dense IoT \cite{BGT_my}. This motivates lightweight, distributed methods that reason under uncertainty without centralized orchestration or heavy data collection. \emph{Epistemology—the study of knowledge distinguishing truth from belief} \cite{epistemology_stanford_Encyclopedia}—offers a foundation for decision-making with incomplete information. In Bayesian game settings, epistemic primitives model players’ beliefs about play and higher-order beliefs about opponents’ beliefs. We leverage Bayesian epistemology to design a resource-allocation framework that prioritizes intelligence-based inference over data-hungry learning, making it suitable for low-power and-memory devices operating under real-time dynamics.

The primary contribution of this paper is the first-ever epistemic Bayesian game framework for distributed IoT uplink power control under incomplete CSI that operates while suppressing interference among multiple uplink channels from distributed devices transmitting simultaneously. We first cast the problem as an imperfect but complete game with a transparent, step-by-step execution diagram, then extend it to a blind equilibrium-seeking mechanism under resource unawareness via inter- and intra-epistemic belief updates that replace exhaustive expected-utility tables. To extract richer structure from uncertainty, we augment utilities with higher-order statistical moments rather than relying on expectation alone, yielding tighter power regulation and improved coverage probability in dense interference. The resulting solver achieves a substantially reduced per-round computational burden, enabling lightweight, on-device operation on inexpensive IoT processors with minimal messaging overhead and fast adaptation to mobility. 

The paper is organized as follows: Section~\ref{Sec: system_model_problem} formulates the problem on the specified system model. Section~\ref{Sec: Game_formation} forms a baseline imperfect-information game and characterizes equilibrium. Section~\ref{Sec: Propsed_epistemology_work} details the proposed epistemic approach, and Section~\ref{Sec: results_and_discussion} reports results. Section~\ref{Sec: conclusion} concludes, and Section~\ref{Sec: appendix} provides appendices.\vspace{-2mm}

\section{System Model and Problem Statement} \label{Sec: system_model_problem}
\begin{figure}[t]
\centering
\includegraphics[width=0.99\columnwidth]{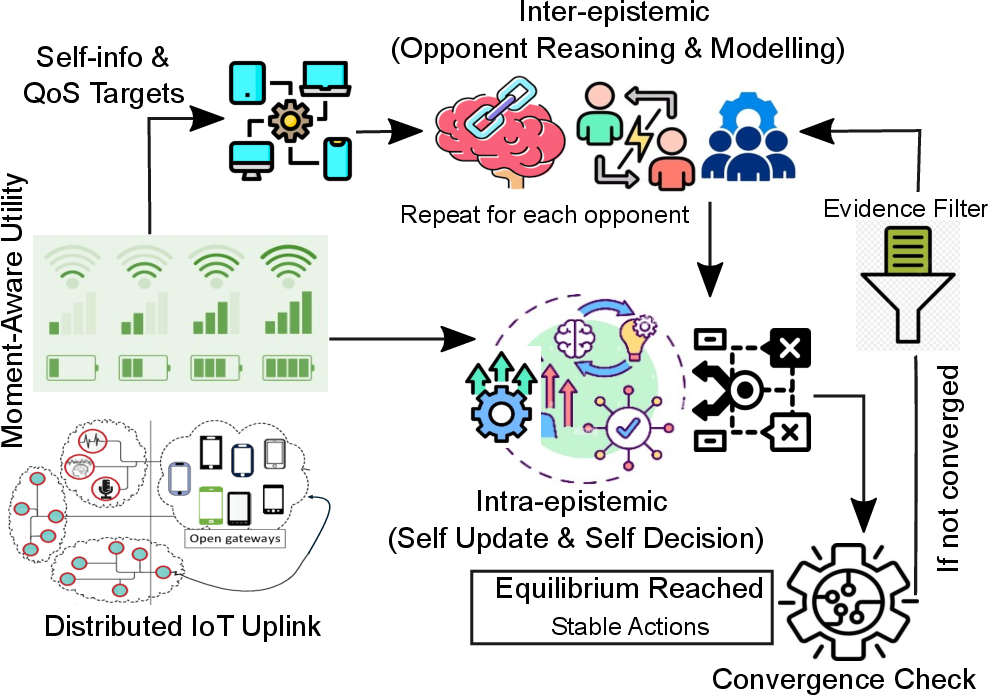}
\centering
\vspace{-5mm}
\caption{Conceptual representation of Epistemic Bayesian Game Model with inter and intra epistemic decision-making approach.}
\vspace{-5mm}
\label{Fig: Epistemic_BGT_System_model}
\end{figure}

During uplink (\figurename~\ref{Fig: Epistemic_BGT_System_model}), IoT devices $i \in \mathcal{N}$ transmit sensed and actuated data randomly to the connected gateway through the allocated orthogonal space-time-frequency spreading (STFS) space. Then, the received signal $y_{i}(t)$ at the gateway 
can be mathematically formulated as \eqref{Eq: rx_sig}. 
\begin{align}\label{Eq: rx_sig}
    y_{i}(t) = g_{i}(t) x_{i}(t) + \sum_{j=1, j\neq i}^{N-1} g_{j}(t) x_{j}(t) + n_{i}(t)
\end{align}
where $x_{i}(t)\in\mathbb{C}$ is the desired transmit signal, $g_{i}(t)\in\mathbb{C}$ is the corresponding complex fading coefficient, the summation captures multiuser interference from devices $j \in \mathcal{N}\setminus i$, and $n_{i}(t)\sim \mathcal{N}(0,\sigma_{n}^{2})$ is zero-mean additive white Gaussian noise. Although nodes occupy independent STFS blocks, realistic propagation—scattering, dispersion and reflection from mobile transceivers and surrounding objects—broadens delay spread and shortens coherence time at higher carrier frequencies, causing partial loss of orthogonality and collisions at the receiver. Thus, the desired stream $i$ is interfered by neighbors $j \in \mathcal{N}\setminus i$.


We evaluate uplink quality of service using the lower bound on the normalized achievable rate over bandwidth $B$ (small-scale fading), i.e., the channel throughput,
\begin{align}\label{Eq:cap_x}
    c_{i} = B\log_2\bigg(1+ \frac{|{g}_i{x}_i|^2}{\sum_{j \in \mathcal{N} \setminus i}|{g}_j{x}_j|^2+\sigma_{n,i}^2}\bigg)
\end{align}
where the SINR of node $i$ is $\gamma_{i}=\frac{|{g}_i{x}_i|^2}{\sum_{j=1, j\ne i}^{N-1}|{g}_j{x}_j|^2+\sigma_{n,i}^2}$ and $\sigma_{n,i}^2$ is the noise power. The transmit signal from desired node $i$ is generated as $x_{i}(t) = \sqrt{p_{i}(t)}\cdot s_{i}(t)$ where the zero-mean data symbol vector $s_{i}(t)$ through a selected modulation scheme and energized by $p_{i}(t) \in \mathbf{P}$ power strength. Therefore, the channel throughput can be re-expressed in terms of power $p_{i}$ as
\begin{align}\label{Eq:cap_p}
    c_{i}& = B\log_2\bigg(1+ \frac{|{g}_i|^2{p}_i}{\sum_{j \in \mathcal{N} \setminus i}|{g}_j|^2{p}_j+\sigma_{n,i}^2}\bigg)
\end{align}
In the uplink, each device can obtain self-CSI via pilot-based estimation, but faces two coupled challenges: (i) estimating neighbors’ CSI and (ii) minimizing transmit power while meeting a throughput target $C_{i}^{th}$. We pose the power-minimization problem as
\begin{align}\label{Eq: P_min_optimization}
\underset{p_{i} \in \mathbf{P}}{\mbox{minimize}}~~ \sum_{i=1}^{N} p_{i}(g_{i}) \quad \text{s.t.}~~ c_{i} \geq C_{i}^{th}; \quad \forall i \in \mathcal{N}.
\end{align}
Transmit-power optimization with incomplete information is NP-hard and combinatorial even under linear constraints \cite{NP_hard_complex_RA}, rendering exhaustive search impractical for low-complexity IoT nodes. Moreover, uplink resource allocation should be realized in a distributed manner rather than at a central gateway, avoiding repetitive physical uplink control channel (PUCCH) signaling and carrier switching overhead in transceiver packets \cite{PUCCH_IoT}. 
Consequently, devices compete to select minimal firing power while satisfying $c_i\!\ge\!C_{i}^{th}$, thereby enhancing network capacity under realistic interference and uncertainty.


\section{Game Formation} \label{Sec: Game_formation}
We model distributed uplink power selection as a non-cooperative game with incomplete information, where each IoT node chooses a transmit-power action to improve gateway quality of service (QoS) while minimizing collision strength. Devices act independently—distributed across the network and unaware of neighbors’ CSI—which motivates a game-theoretic formulation.
\begin{align}
        \mathcal{G} \triangleq \Big\langle \mathcal{N}, \mathcal{T}, \phi_{\mathcal{T}}, \{ \mathcal{S}_{i}, \mathcal{U}_{i} \}_{i \in \mathcal{N}} \Big\rangle
        \label{Eq: Game_model_common}
\end{align}
Here, players are the IoT devices $i \in \mathcal{N}$. Each player occupies a state of nature (type) $t_i \in \mathcal{T}$; equivalently, with individual type spaces $\mathcal{T}_i$, the joint type space satisfies $\mathcal{T}=\times_{i \in \mathcal{N}} \mathcal{T}_i$ (i.e., $\{\times_{i \in \mathcal{N}} t_i\}=\mathcal{T}$). Types are drawn from a common prior $\phi_{\mathcal{T}}$ shared by all players, with normalization $\int_{-\infty}^{\infty} \phi_{\mathcal{T}} \, dt_i~=~1;\ \forall i \in \mathcal{N}$. Player $i$ selects an action from a finite strategy set $\mathcal{S}_i$, and the joint strategy space is $\mathcal{S}=\times_{i \in \mathcal{N}} \mathcal{S}_i$. The opponents’ strategy profile is $\mathcal{S}_{-i}=\{ \mathcal{S}_1,\dots,\mathcal{S}_{i-1},\mathcal{S}_{i+1},\dots,\mathcal{S}_N \}$. Given type $t_i$ and strategy profile, player $i$ accrues utility $\mathcal{U}_i$, with $\mathcal{U}=\{c_1,\dots,c_N\}$ and $c_i:\mathcal{S}\times\mathcal{T}\rightarrow\mathbb{R}$ defined by the throughput in \eqref{Eq:cap_p}. In accordance with the resource-allocation statement \eqref{Eq: P_min_optimization}, devices act as players and choose transmit-power vectors via the mapping $\mathcal{S}_i \rightarrow p_i$. The incomplete channel-gain vector among nodes is mapped to the player types $(g_i \leftarrow \mathcal{T})$, with small-scale fading captured by the prior distribution $\phi_{\mathcal{T}}$. This formulation captures the two core hurdles: (i) players cannot foresee opponents’ future actions, yielding an \emph{imperfect} game; and (ii) lack of neighboring CSI induces \emph{incomplete information}. Together, these features necessitate a Bayesian (type-dependent) treatment of uplink power control under interference.\vspace{-0mm}
\begin{figure}[t]
\centering
\vspace{-2mm}
\includegraphics[width=0.8\columnwidth, trim={0mm 1mm 1mm 3mm},clip]{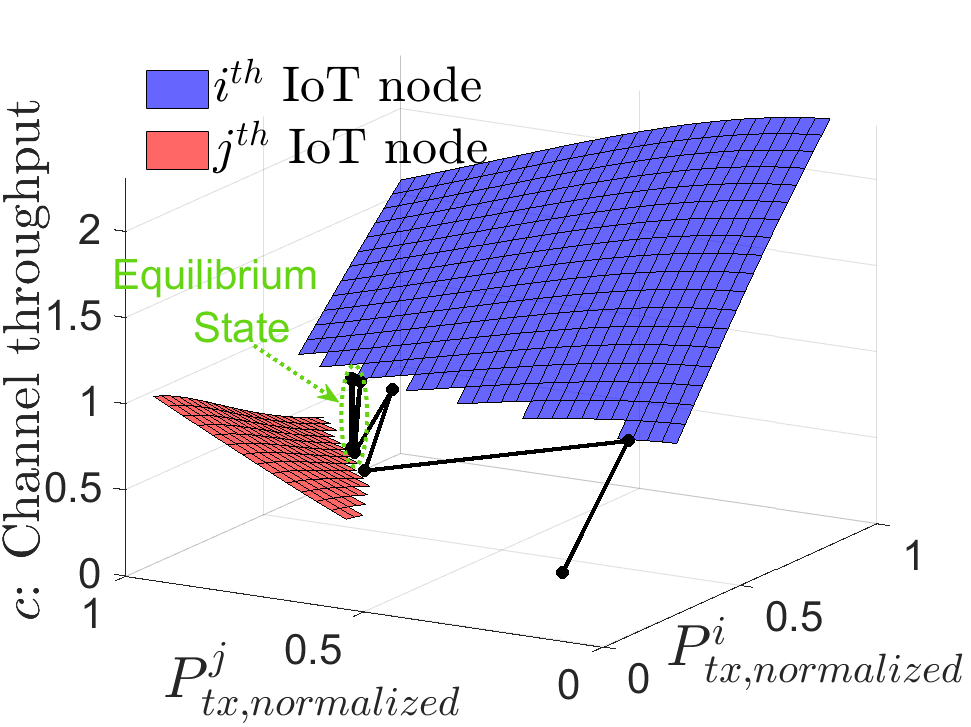} 
\vspace{-2mm}
\caption{Visualization ($3D$ Schematic) of normalized transmit power strategy convergence pattern toward the equilibrium for $2$ IoT nodes, named $i$ and $j$, with CSI $\left| g_{i} \right| =0.2$ and $\left| g_{j} \right|=0.1$ satisfying the given channel throughput threshold values (linear) $\gamma_{i}= 0.8$  and $\gamma_{j}=0.4$. 
}
\label{Fig: S_K_zeta_0_4}
\vspace{-6mm}
\end{figure}
\emph{Imperfect Game:} In the framework, we first study an \emph{imperfect} game in which all IoT devices possess perfect CSI—i.e., a complete-information game—yet cannot observe or predict the post-transmit power selections of other nodes. Under the model in \eqref{Eq: Game_model_common}, every player knows the opponents’ instantaneous type space, but action choices are unrevealed at the time of decision, creating strategic uncertainty. 
Here, each player’s payoff is the channel throughput, i.e., $\mathcal{U}_{i}\!\leftarrow\!c_{i}$. For $p_{i}\!\in\!\mathcal{S}_{i}$, the function $c_{i}(\mathbf{p})$ is continuous in the joint action $\mathbf{p}$ and concave in $p_{i}$ for any fixed opponent strategy profile $\times_{j \in \mathcal{N}\setminus i}\mathcal{S}_{j}$. Hence a (Nash) equilibrium of the concave $n$-person game is a point $\mathbf{p}^{*}$ such that, for all $i\in\mathcal{N}$, $c_{i}(\mathbf{p}^{*}) \;=\; \max_{x_{i}} \{ c_{i}\big(p_{1}^{*},\ldots,p_{i-1}^{*},x_{i},p_{i+1}^{*},\ldots,p_{n}^{*}\big) \ \big|\ (p_{1}^{*},\ldots,x_{i},\ldots, \newline p_{n}^{*}) \in\mathbb{R} \}.$ At $\mathbf{p}^{*}$ no player can improve her payoff by a unilateral deviation, by \emph{the property of Existence \& Uniqueness}
(see proof in \cite{exist_n_uniquness_NE_proof}), then the graphical illustration
is provided in \figurename~\ref{Fig: S_K_zeta_0_4}, where equilibrium appears as the intersection of the contour objectives \emph{(two-dimensional)} projected from players’ utility surfaces. 
Operationally, each IoT node computes a best response given opponents’ CSI and admissible strategies and eliminates strictly dominated actions \cite{IESDS_strictly_dominated_strategy}, iterating toward the equilibrium (global optimum in this concave setting), even with stablizing for heterogeneous throughput targets $\{C_{i}^{th},C_{j}^{th}\}$.
\vspace{-2mm}
\section{Epistemology for Proposed Framework} \label{Sec: Propsed_epistemology_work}
\vspace{-2mm}
The uplink resource assignment cannot be solved as a purely imperfect-information game: direct iteration to the optimum is blocked by incomplete knowledge of neighboring CSI. Accordingly, IoT players $i\in\mathcal{N}$ are assumed rational and never select strictly dominated actions with respect to opponents’ strategies $\mathcal{S}_{-i}$. The best rational response based on beliefs $s^{*}\!\in\!\mathcal{S}_{i}$, with $Pr(s_{i}\!\leftarrow\! t_{i})\!\in\!\phi_{\mathcal{T}}$ for all $s\!\in\!\mathcal{S}_{i}$, is
\begin{align} \label{Eq: rational_best_response}
    \sum_{s_{-i}\in \mathcal{S}_{-i}} Pr(s_{-i}).U_{i}\left( s^{*},s_{-i} \right) \geq \sum_{s_{-i}\in \mathcal{S}_{-i}} Pr(s_{-i}).U_{i}\left( s,s_{-i} \right)
\end{align}
which places the desired node $i$ in a probabilistic utility space over interfering neighbors and motivates an expectation-based mechanism.\vspace{-3mm}

\subsection{Utility Expectation}\label{Sec: Utility_expectation}
The payoff function $c_{i}$ of the desired $i$th node is a function of the SINR in terms of transmit power $p \in \textbf{P}$ as $\gamma_{i}=\frac{|{g}_i|^2{p}_i}{\sum_{j=1, j\ne i}^{N-1}|{g}_j|^2{p}_j + \sigma_{n,i}^2}$. Alternatively, we can simplify the $\gamma_{i}$ given by, $\gamma_{i} \coloneqq {X}/{[Y+\eta]}$, where $X \sim exp(\lambda); \quad x \geq 0$ is exponentially distributed with rate parameter $\lambda=\frac{1}{2\sigma^2}$ which is derived through the Rayleigh distribution, $|g_{i}| \sim Rayl(\sigma)$. And $Y$
is the summation of exponential random variables $|g_{j}|^2 \sim exp(\lambda); \quad |g_{j}| \geq 0$ that are multiplied by scalar power vector $p_{j} \in \boldsymbol{p}$, $\forall{j} \in \mathcal{N}\setminus i$. Therefore, we can derive the approximate Gamma distribution for $Y \sim \Gamma(\hat{\alpha}, \hat{\theta}); \quad y \geq 0$, \eqref{Eq: gamma_distriution_estimated} for the interference term, including the estimated shape parameter $\hat{\alpha}=\frac{\left(\sum_{j=1}^{N-1} p_{j}\right)^2}{\sum_{j=1}^{N-1}p_{j}^2}$ and the scale parameter $\hat{\theta} =\frac{\sum_{j=1}^{N-1} p_{j}^2}{\lambda \sum_{j=1}^{N-1}p_{j}}$, (proof: Appendix \ref{appendix: estimated_Gamma_distribution}). And, the constant $\eta$ is considered as the summation of the desired signal AWGN noise power
and a supportive scalar term $(\eta = |g_{i}|^2p_{i}^{l} + \sigma^2)$ representing the desired signal $i\in \mathcal{N}$ received power term when acting as an interferer for an arbitrary desired opponent IoT node $j^{'}\in \mathcal{N}$ in the inter-epistemic belief hierarchy in subsequent derivations \eqref{Eq: inter_epist_H}, \eqref{Eq: inter_epist_E}. In contrast, $(\eta = \sigma^2)$ is the AWGN noise power only for the intra-epistemic stages when the rational beliefs-taking player is the desired one $i\in \mathcal{N}$, which is explained in \eqref{Eq: intra_epist_H}, \eqref{Eq: intra_epist_E}. Then, the expected utility $\bar{\gamma_{i}}$ of the desired IoT device $i$ is simplified 
using (proof: Appendix \ref{appendix: expected_utility}), and insert \eqref{Eq: expected_Y_term} into \eqref{Eq: utility_bar}.
\vspace{-3mm}
\begin{align} 
    \begin{split}\label{Eq: expected_SINR_gamma}
         \bar{\gamma_{i}} = \frac{E[X]}{E[Y]+\eta} \left( \sum_{n=0}^{\infty} \frac{(-1)^n E\left[ \left( Y-E[Y] \right)^n \right]}{\left( E[Y]+\eta \right)^n} \right)
    \end{split}
\end{align}
Our epistemic design identifies opponents’ latent states and builds a conditional belief hierarchy—transforming hypotheses into evidence—so that node $i$ selects $s^{*}$ via belief updates rather than exhaustive enumeration, preserving all statistical structure and the inter-/intra-epistemic roles of $\eta$.\vspace{-3mm}
\subsection{Conditional Belief}\label{Sec: Conditional_beliefs}
\vspace{-1mm}
We now quantify how assumptions on players’ beliefs and rationality shape strategic choices in the epistemic setting. \figurename~\ref{Fig: Epistemic_BGT_System_model} summarizes the algorithm executed by the desired Internet of Things (IoT) device $i$ along two coupled iterative axes: \emph{inter-epistemic} and \emph{intra-epistemic} updates. Device $i$, knowing only its self-channel gain $g_{i}$, takes an initial action $p_{i}^{(l=0,m=0)}$ to seed the belief structure. It then “enters” the interference pool and hypothesizes over an immediate opponent $j^{'}$, treating $j^{'}$ as the desired transmitter, yielding the hypothesis set in \eqref{Eq: inter_epist_H}. Among these hypotheses, the one that satisfies $j^{'}$’s throughput threshold with the highest belief weight becomes evidence in \eqref{Eq: inter_epist_E} at state $(l,m)$ and determines the best instantaneous action $(p_{j^{'}}^{m})^{*}$. This constitutes an \emph{inter-epistemic} update; node $i$ repeats it for all competing interferers. For fixed intra stage $l$, the inter-epistemic transitions from $(m\!-\!1)$ to $m$ are
\begin{align}
    \mathcal{H}_{(j^{'}, inter)}^{(l,m)} &= \left\{ \mathbb{E} \left[ \gamma_{j^{'}} \left( p_{j^{'}}^{m} \right) \Big| \left\langle |g_{i}|, \left\{p_{i}^{l}, p_{j \in \mathcal{N}\setminus \{i,j^{'}\}}^{(m-1)} \right\} \right\rangle \right] \right\}_{p_{j^{'}}^{m} \in \textbf{P}} \label{Eq: inter_epist_H}\\
    \mathcal{E}_{(j^{'}, inter)}^{(l,m)} &: \exists \mathbb{E} \left[ \gamma_{j^{'}} \left( p_{j^{'}}^{m} \right)^{*} \Big| \left\langle |g_{i}|, \left\{p_{i}^{l}, p_{j \in \mathcal{N}\setminus \{i,j^{'}\}}^{(m-1)} \right\} \right\rangle \right] \geq \gamma_{j^{'}}^{th} \label{Eq: inter_epist_E}
\end{align}
\vspace{-1mm}
After processing all competitors ($m \rightarrow \mathcal{M}_{l}$) at stage $l$, device $i$ performs an \emph{intra-epistemic} self-update—conditioning on the newly filtered evidence from the inter step—and advances from $l$ to $l\!+\!1$ for the given $m$:
\begin{align}
    \mathcal{H}_{(i, intra)}^{(l+1,m)} &= \left\{ \mathbb{E} \left[ \gamma_{i} \left( p_{i}^{(l+1)} \right) \Big| \left\langle |g_{i}|, \left\{ p_{j \in \mathcal{N}\setminus \{i\}}^{(l,m)} \right\} \right\rangle \right] \right\}_{p_{i}^{(l+1)} \in \textbf{P}} \label{Eq: intra_epist_H}\\
    \mathcal{E}_{(j^{'}, intra)}^{(l+1,m)} &: \exists \mathbb{E} \left[ \gamma_{i} \left( p_{i}^{(l+1)} \right)^{*} \Big| \left\langle |g_{i}|, \left\{ p_{j \in \mathcal{N}\setminus \{i\}}^{(l,m)} \right\} \right\rangle \right]  \geq \gamma_{i}^{th} \label{Eq: intra_epist_E}
\end{align}
Iterations proceed until the global equilibrium $(l^{*},m^{*}) \in (\mathcal{L},\mathcal{M})$ where no IoT player has incentive to deviate. In the epistemic-probability model, each stage encodes both players’ choices and each player’s beliefs about others’ choices; stage transitions are driven by maximizing conditional belief mass—i.e., hypotheses at the current stage are conditioned on evidence distilled from the immediate prior stage. Formally, the belief of player $i$ about event $E$ conditioned on $F$ is $B^{F}_{i}\left( E \right) = \left\{ w: Max_{\geq i} \left[ F \cap \Psi_{i}(w) \right] \subseteq E \right\}$ where $\Psi_{i}(w)$ is a partition of the state space containing $w=|g_{i}|$, and the inter-epistemic events relate as $F_{\text{inter}}=\mathcal{E}_{j^{'}}^{(l,m-1)}$ (current conditioning event) and $E_{\text{inter}}=\mathcal{H}_{j^{'}}^{(l,m)}$ (current hypothesis). Analogously, for intra-epistemic updates $F_{\text{intra}}=\mathcal{E}_{i}^{(l,m)}$ and $E_{\text{intra}}=\mathcal{H}_{i}^{(l+1,m)}$ are the consecutive conditioning–hypothesis pair. {Operationally,} each node maintains and updates an independent strategic map of competitors using only self-CSI and the prior distribution over others. Conditioning hypotheses on stage-wise evidence steers all players toward the global equilibrium in the expected domain, \emph{without} constructing exhaustive expected-utility tables. This belief-driven reduction is the key to the framework’s substantial computational savings.\vspace{-3mm}


\subsection{Higher order statistics}\label{Sec: HoS}
The probabilistic utility space induced by the prior $\phi_{\mathcal{T}}\!\sim\!\mathrm{Rayl}(\sigma)$ is inherently skewed with non-negligible tails. Relying on expectation alone can therefore obscure decisive structure in the payoff distribution. We extend the epistemic update to \emph{higher-order} statistics of the strategic hypothesis space. Let $m_{k,Z}$ and $\bar{m}_{k,Z}$ denote the $k$-th raw and central moments of a random variable $Z$, respectively. Using \eqref{Eq: utility_bar} and \eqref{Eq: k_th_order_moment_of_Y}, the $k$-th raw moment of the SINR payoff $\gamma$ admits (proof in Appendix~\ref{appendix: higher_order_moments})
\begin{align}
    \begin{split}
        &m_{k,\gamma} = \frac{m_{k, X}}{\left( E[Y]+\eta \right)^k}\sum_{n=0}^{\infty} (-1)^n \frac{(k+n-1)!}{(k-1)!.n!}\frac{\bar{m}_{n,Y}}{\left(E[Y]+\eta \right)^n};\\
    &m_{k, X} = E[X^{k}] = \frac{k!}{\lambda^{k}}; \quad m_{n,Y} = \hat{\theta}^{n} \prod_{\kappa=1}^{n}(\hat{\alpha}+\kappa-1), 
    \end{split}
\end{align} 
\vspace{-1mm}
where $X$ and $Y$ follow the exponential–Gamma construction established earlier and $\eta$ is defined by the inter-/intra-epistemic roles. Incorporating variance, skewness, and higher moments into the belief hierarchy equips each device with richer uncertainty descriptors, enabling sharper power updates and fine-grained, opponent-aware tuning of optimal transmit power—beyond what expectation-only policies can capture.
\vspace{-5mm}
\subsection{Complexity}\label{Sec: complexity}
We take one payoff expectation as $\mathcal{O}(1)$ (baseline). A traditional Bayesian solver must (i) precompute the expected-utility space at cost $\mathcal{O}\!\left(N\,S^{TN}\,T^{N}\right),$ since there are $S^{TN}$ strategy profiles across $T^{N}$ type profiles, and then (ii) perform an exhaustive equilibrium search over an NP-hard space with cost $\mathcal{O}\!\left(N\,S^{T}\,S^{TN}\right)=\mathcal{O}\!\left(N\,S^{T(1+N)}\right),$ yielding total complexity $\mathcal{O}\!\left(N\,S^{TN}\,T^{N}\;+\;N\,S^{T(1+N)}\right).$ Alternatively, best-response over the full expected table requires $\mathcal{O}\!\left(N\,S^{T(N+1)}\,T^{N}\right)$ for $S^{NT}$ cycles. In contrast, our epistemic scheme computes the inter-belief hierarchy in \eqref{Eq: inter_epist_H}–\eqref{Eq: inter_epist_E} with cost $\mathcal{O}\!\left(N\,S^{N}\right)$ for $S^{N}$ strategy profiles per desired node, followed by intra-belief convergence \eqref{Eq: intra_epist_H}–\eqref{Eq: intra_epist_E} over at most $S^{\bar{N}}$ cycles (with $\bar{N}\!\le\!N$ leftover nodes). Thus a single allocation realization over $N$ devices is upper-bounded by $\mathcal{O}\!\left(N\,\bar{N}\,S^{N}\,S^{\bar{N}}\right)\;\leq\;\mathcal{O}\!\left(N^{2}\,S^{2N}\right).$

\vspace{-2mm}
\section{Results and Discussion} \label{Sec: results_and_discussion}
\begin{figure}[!t]
\centering
\includegraphics[width=0.9\columnwidth, trim={0mm 1mm 1mm 5mm},clip]{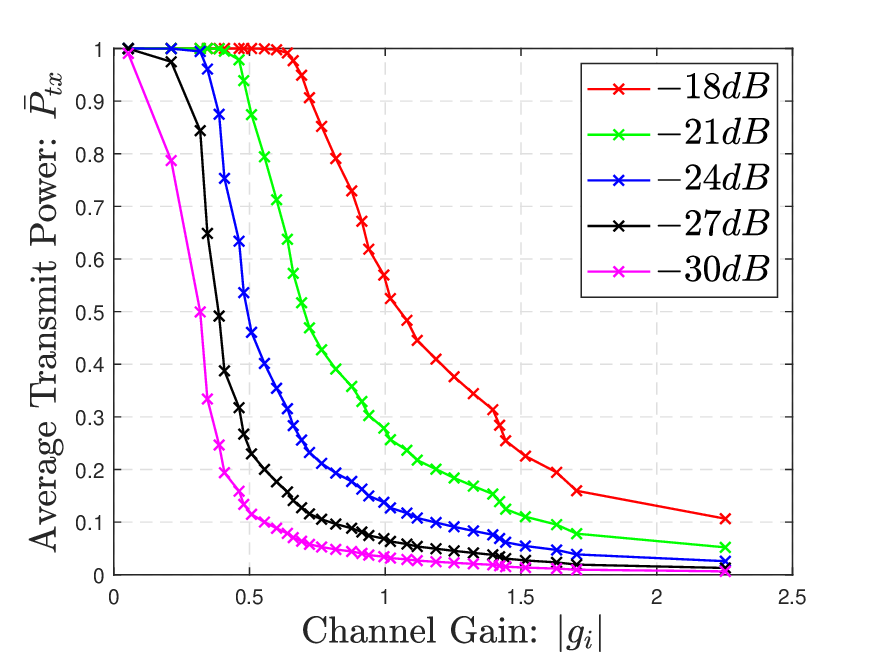}
\vspace{-4mm}
\caption{Variation of Normalized average transmit power $(\bar{P}_{tx,i} \text{ Vs } \left| g_{i} \right|)$ of the desired IoT node $i$, against Rayleigh channel gain with unity average power for given SINR threshold $\gamma_{i}^{th}$ values.}
\vspace{-7mm}
\label{rate_snr}
\end{figure}
We validate the proposed epistemology-based resource-allocation framework using two metrics: (i) average transmit power and (ii) coverage capability for an arbitrary IoT node. We also assess network reliability beyond expectation-only utilities by identifying potential gateway drops caused by collisions with interfering neighbors and by the relative strength of their transmissions. Unless stated otherwise, simulations consider an IoT network with $100$ nodes \emph{(not limited to)}, Rayleigh variance $\sigma^{2}=1$ (w.l.o.g.), noise power $-120\,\mathrm{dB}$, and an upper SINR margin $\gamma^{th}_{i}=-20\,\mathrm{dB}$.

\begin{figure}[t]
\centering
\includegraphics[width=0.9\columnwidth, trim={0mm 1mm 1mm 5mm},clip]{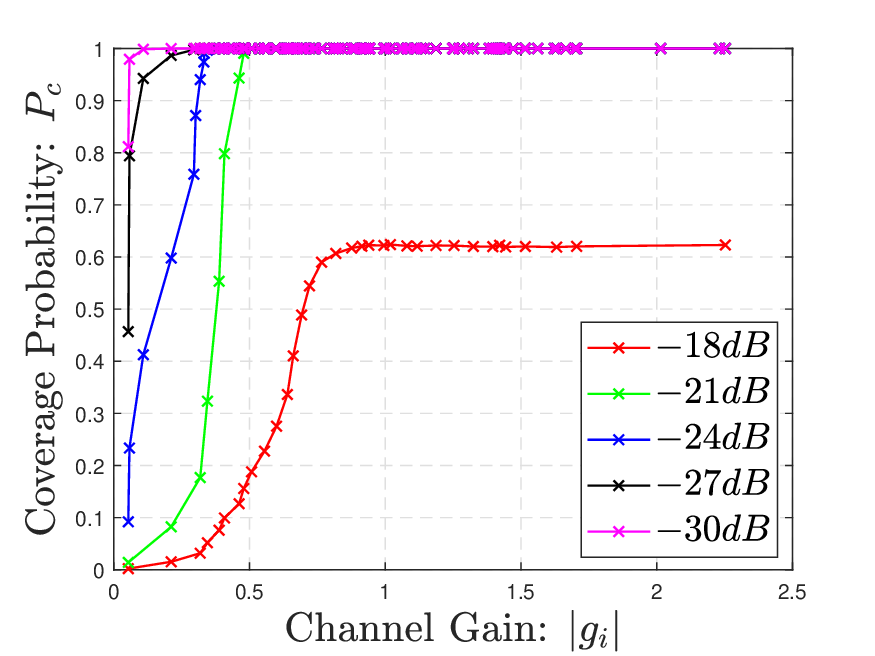} 
\vspace{-4mm}
\caption{Variation of Normalized coverage probability $(\Pr_{cov} \text{ Vs } \left| g_{i} \right|)$, of the desired IoT node $i$, against Rayleigh channel gain with unity average power for given SINR threshold $\gamma_{i}^{th}$ values.
}
\label{Fig: Avg_Ptx_and_Pcov_vs_channel_gain}
\vspace{-6mm}
\end{figure}
As shown in \figurename~\ref{rate_snr}, the required uplink transmit power decreases with increasing channel gain, a key property of the protocol. Minimal firing strengths vary with the SINR target $\gamma^{th}_{i}$, reflecting heterogeneous resource demands: for example, a device with $\lvert g_{i}\rvert=0.5$ requires approximately $10\%\,p^{\max}$ to meet $\gamma^{th}_{i}=-30\,\mathrm{dB}$ (low-rate, narrowband), whereas devices pursuing higher data rates with $\gamma^{th}_{i}=-18\,\mathrm{dB}$ require about $100\%\,p^{\max}$ while achieving stronger QoS. Physically, this set of curves refers to the fact that better channels require less transmit power to meet SINR/throughput targets, while stricter targets (higher data rates) demand near-maximum power—so the scheme adaptively sets each node’s power to the minimum needed to satisfy QoS, thereby reducing interference and energy use.

\figurename~\ref{Fig: Avg_Ptx_and_Pcov_vs_channel_gain} shows the coverage probability \(\Pr_{cov}\) of node \(i\) versus the channel gain \(\lvert g_i\rvert\) for diverse SINR thresholds: low \(\gamma_i^{th}\) expands the achievable coverage region at high channel gains, i.e., nodes targeting low data rates maintain high \(\Pr_{cov}\) even under a low link budget. Notably, with \(\lvert g_i\rvert=1\) and \(\gamma_i^{th}=-18\,\mathrm{dB}\), the node attains \(\approx 60\%\) maximum coverage using \(\approx 55\%\,p_{\max}\); mid-rate devices with \(\gamma_i^{th}=-27\,\mathrm{dB}\) reach full coverage \((\Pr_{cov}=1)\) with \(<0.1\%\,p_{\max}\) under poor link conditions. Physically, favorable channel conditions and loose SINR targets reduce the power required for reliable decoding, whereas tight targets demand high transmit power—so the policy selects the minimum transmit power that satisfies reliability, thereby minimizing interference and energy use. In traditional Bayesian game theory, players rely on expected utility to locate equilibrium; while averaging over competitors is reasonable under symmetric, well-behaved priors \(\phi_{\mathcal{T}}\), it breaks down for practical distributions with asymmetry and heavy tails. Consequently, we exploit the first four moments—expectation, variance, skewness, and excess kurtosis—of the hypothesis space to capture salient distributional structure and enhance the effectiveness of the proposed resource-allocation framework.

\begin{figure}[!t]
\centering
\includegraphics[width=0.9\columnwidth, trim={0mm 1mm 1mm 5mm},clip]{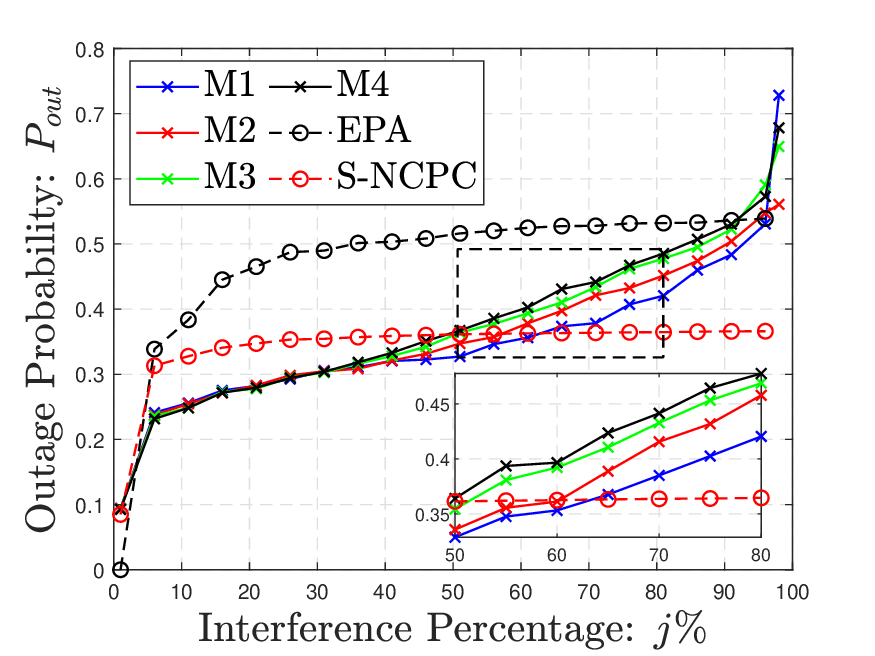}
\vspace{-4mm}
\caption{Performance comparison of coverage probability $(\Pr_{out} \text{ Vs } j\% )$ of the IoT network $i \in \mathcal{N}$, against the interference strength $j\%$ with Rayleigh channel gain with unity average power for the set of HoS $M_{k, \gamma_{i}}$ values of SINR $\gamma_{i}$ bounded by $\gamma_{i}^{th}=-20\,\mathrm{dB}$ then the baselines of EPA and S-NCPC.}
\vspace{-3mm}
\label{Fig: rate_snr2}
\end{figure}

\begin{figure}[t]
\centering
\includegraphics[width=0.9\columnwidth, trim={0mm 1mm 1mm 7mm},clip]{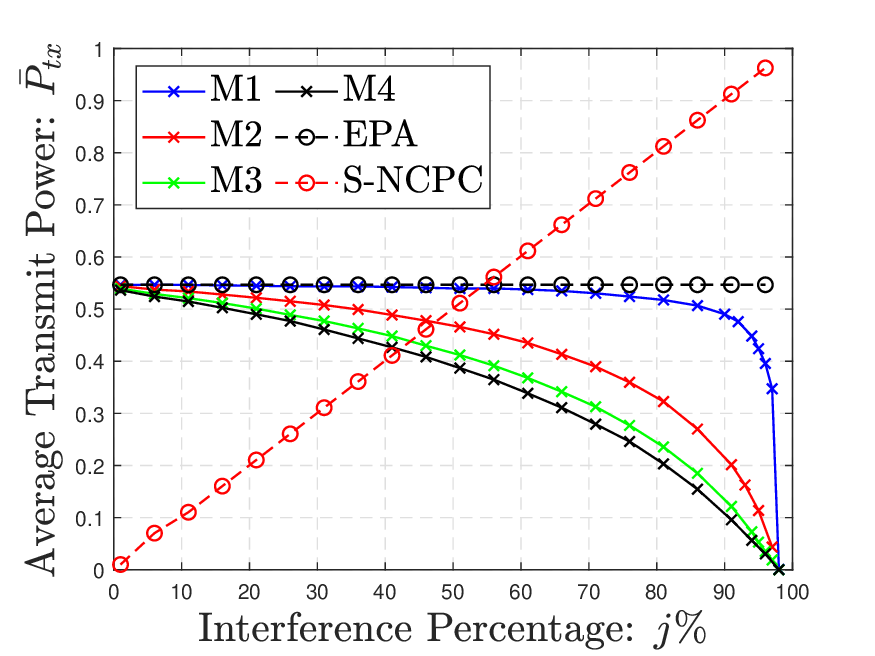}
\vspace{-3mm}
\caption{Performance of Normalized average transmit power $(\bar{P}_{tx} \text{ Vs } j\% )$, of the IoT network $i \in \mathcal{N}$, against the interference strength $j\%$ with Rayleigh channel gain with unity average power for the set of HoS $M_{k, \gamma_{i}}$ values of SINR $\gamma_{i}$ bounded by $\gamma_{i}^{th}=-20\,\mathrm{dB}$ then the baselines of EPA and S-NCPC.}
\label{Fig: P_avg_nw_and_P_out_nw_vs_interf_percentage}
\vspace{-7mm}
\end{figure}

\figurename~\ref{Fig: rate_snr2} reports network outage probability \(\Pr_{out}\) versus the percentage of interfering users \(j\%\). Outages increase monotonically with \(j\%\), reflecting more collisions at the gateway. For low interference (\(j\%\!\le\!30\)), all utility–moment policies and baselines are nearly indistinguishable; for \(50\!\le\!j\%\!\le\!90\), the moment policies separate slightly: the first moment \(M_{1,\gamma_i}\) attains the lowest \(\Pr_{out}\), the fourth moment \(M_{4,\gamma_i}\) the highest, with a negligible gap \((\delta \Pr_{out})^{\max}\!\sim\!0.05\). Relative to benchmarks, Equal Power Allocation (EPA) and Stochastic–Non-Cooperative Power Control (S–NCPC)—the latter optimizing the desired (self) transmit power via a statistical–interference power domain—exhibit higher \(\Pr_{out}\) in this moderate–high contention region, demonstrating the reliability advantage of belief-driven control. Consequently, as concurrency rises, aggregate interference and decoding failures grow; expectation-based epistemic control tracks the central tendency of interference and is marginally more reliable than kurtosis-focused control, while both surpass EPA and S–NCPC when contention is high.

\figurename~\ref{Fig: P_avg_nw_and_P_out_nw_vs_interf_percentage} examines average network transmit power versus \(j\%\). Power decreases as interferers increase, matching interactive, game-theoretic adaptation (\emph{low loud for low conflicts}). At \(80\%\) interferers, the expectation policy \(M_{1,\gamma_i}\) uses \(\sim 52\%\,p^{\max}\), whereas the fourth-moment policy \(M_{4,\gamma_i}\) reduces this to \(\sim 20\%\,p^{\max}\) under identical constraints; \(M_{3,\gamma_i}\) lies between these. Although S–NCPC can consume less transmit power than the proposed approach at low interference, its \(\Pr_{out}\) is higher (\figurename~\ref{Fig: rate_snr2}); as \(j\%\) grows, the relationship reverses—our moment-aware policies require less power than S–NCPC and clearly undercut EPA \emph{while} maintaining target reliability. Towards this end, higher contention amplifies interference variability; skewness- and kurtosis-aware policies hedge against tail events, enabling the network to meet reliability with substantially lower average power than expectation-only tuning and baseline strategies. As a result, the epistemology-inspired protocol lowers transmit-power demand in a way compatible with acceptable transmission capabilities, outperforming EPA and S–NCPC across practical high-interference regimes.
\section{Conclusion} \label{Sec: conclusion} 
\vspace{-0mm}
We introduced an epistemology-inspired Bayesian game framework for distributed IoT uplink power control under incomplete CSI, replacing exhaustive expected-utility tables with inter-/intra-epistemic belief updates over an exponential–Gamma SINR model and augmenting utilities with higher-order moments (expectation, variance, skewness, excess kurtosis). The design achieves lightweight, on-device optimization with a single-round upper bound of \(O\!\left(N^{2}S^{2N}\right)\), while suppressing simultaneous multi-channel interference. Simulations show precise power regulation and strong reliability, thereby improving energy efficiency and curbing cross-channel collisions. These results establish a principled, computationally lean path to scalable, interference-aware resource allocation for dense IoT; future work will validate the method on empirical channel/interference traces and extend the belief hierarchy to online learning and protocol co-design.
\section{Appendix} \label{Sec: appendix}
\subsection{Estimated Gamma Distribution} \label{appendix: estimated_Gamma_distribution}
Here we estimate the shape and scale parameter of the approximate Gamma distribution relevant to the interference term $Y = \sum_{j=1, j\ne i}^{N-1}x_{j}p_{j}$ where $\quad f_{X}(x_{j}) = \lambda \exp{(-\lambda x_{j})}; \quad x_{j}\geq 0$ in the denominator of \eqref{Eq: expected_SINR_gamma}. Method of moments estimators (MMEs) are utilized to equate sample moments with the population moments ($m_{k}: k^{th}$ order raw moment of random variable $y \in  Y$) to estimate $\hat{\alpha}$ and $\hat{\theta}$ as follows \cite{methods_of_moments_MoM}:
\begin{align} \label{Eq: gamma_dist_k_theta_formulate}
    \begin{split}
        \hat{\alpha} &= \frac{m_{1}^2}{m_{2}-m_{1}^2} = \frac{E[Y]^2}{V[Y]} \text{, and } \hat{\theta} = \frac{m_{2}-m_{1}^2}{m_{1}} = \frac{V[Y]}{E[Y]}\\
        &; E[Y] = \frac{1}{\lambda}\sum_{j=1, j\ne i}^{N-1} p_{j} \text{, and }  V[Y] = \frac{1}{\lambda^2}\sum_{j=1, j\ne i}^{N-1} p_{j}^2
    \end{split}
\end{align}
Then, the Gamma distribution can be formulated by:
\begin{align} \label{Eq: gamma_distriution_estimated}
    Y \sim \Gamma \left(\frac{\left(\sum_{j=1}^{N-1} p_{j}\right)^2}{\sum_{j=1}^{N-1}p_{j}^2}, \frac{\sum_{j=1}^{N-1} p_{j}^2}{\lambda \sum_{j=1}^{N-1}p_{j}} \right)
\end{align}

\subsection{Derivation of expected utility} \label{appendix: expected_utility}
The desired and interference terms are distributed independently in the utility function $\gamma_{i}$ of node $i$.
\begin{align}\label{Eq: utility_bar}
    \bar{\gamma_{i}} &=E\left[ X \right].E\left[ \frac{1}{Y+\eta} \right]; \quad X \indep Y
\end{align}
Let's work on the non-linear mapping function $g(Y)=\frac{1}{Y+\eta}$, 
\begin{align} \label{Eq: Y_term}
    \begin{split}
       g(Y) &\triangleq \frac{1}{E[Y] + (Y+\eta)-E[Y] }; \quad \Bigg|\frac{Y-E[Y]}{E[Y]+\eta} \Bigg| \leq 1 \\
        &= \frac{1}{E[Y]+\eta}\left( 1 - \frac{Y-E[Y]}{E[Y]+\eta} +  \left(\frac{Y-E[Y]}{E[Y]+\eta} \right)^2 - \dots\right)\\
        ;& \quad \text{following to the Maclaurin series expansion}
    \end{split}
\end{align}
Taking the expectation of \eqref{Eq: Y_term},
\begin{align}\label{Eq: expected_Y_term}
    E\left[\frac{1}{Y+\eta}\right] = \frac{1}{E[Y]+\eta} \left( \sum_{n=0}^{\infty} \frac{(-1)^n E\left[ \left( Y-E[Y] \right)^n \right]}{\left( E[Y]+\eta \right)^n} \right)
\end{align}
\vspace{-3mm}
%
\subsection{Derivation Higher order Moments for utility function} \label{appendix: higher_order_moments}
Based on the proof \eqref{Eq: expected_Y_term} the first order moment of the utility function, $\bar{\gamma}_{i}$ explained in appendix \ref{appendix: expected_utility}, we follow the Taylor polynomial through the expectation of interference summation $E[Y]$, as the point of expansion to formulate the higher order statistics of $\gamma_{i}$. 
\begin{align} \label{Eq: Taylor_series}
        g(Y) &= \sum_{n=0}^{\infty} \frac{g^{n}(E[Y])}{n!}\left(Y-E[Y]\right)^{n}
\end{align}
$k^{th}$ order moment for $g(Y)$ is, \(\mapsto E\left[\frac{1}{\left( Y+\eta \right)^k}\right]\)
\begin{align} \label{Eq: k_th_order_moment_of_Y}
    \begin{split}
        %
        m_{k,g(Y)} &= \frac{1}{\left( E[Y]+\eta \right)^k}\sum_{n=0}^{\infty} (-1)^n \frac{(k+n-1)!}{(k-1)!.n!}\frac{\bar{m}_{n,Y}}{\left(E[Y]+\eta \right)^n}\\
        &; \bar{m}_{n,Y} = E\left[ (Y-E[Y])^n \right]
    \end{split}
\end{align}
\vspace{-3mm}
\bibliographystyle{IEEEtran}
\bibliography{IEEEabrv,ref}

\end{document}